\documentclass[12pt]{iopart}
\usepackage{amsfonts,graphicx}
\usepackage{url}

\newcommand{\Pomeron}{\mathbb{P}}

\begin{document}
\title{Factorization in hard diffraction}

\author{J Collins}

\address{%
        Physics Department,
        Penn State University, 
        104 Davey Laboratory,
        University Park PA 16802
        U.S.A.
}
\ead{collins@phys.psu.edu}

\begin{abstract}
In this talk, I reviewed the role of factorization in diffraction hard
scattering.  
\end{abstract}

\section{Introduction}

I start by reviewing one of the complications that produces seemingly
incompatible ways of viewing hard diffraction.  This is that some
people like the target rest frame, where the dipole picture is most at
home, while other people like the Breit frame, where hard-scattering
factorization is the natural point of view.

Then I explain what factorization is, and review some of the
complications, especially as concerns final-state interactions.

\section{Dipoles v.\ pdfs in electroproduction}

There are two distinct communities that differ in the reference frame
they use and in the associated natural intuitions.  The two views have
to be compatible, since we are all discussing the same theory and some
of the same phenomena.  It is important to understand the
compatibility of the two views and the strengths and limitations of
the intuitive ideas, but this is quite non-trivial.  In
electroproduction, the views are:
\begin{enumerate}
\item Target rest frame.  In this frame the virtual photon is of very high
  energy, and it normally dissociates into a (virtual) hadronic system
  long before the interaction with the target.  The intuition comes from
  observing that time dilation implies that the time to cross the proton
  is much less than the formation time of the hadronic component of the
  virtual photon:
  \begin{equation}
     t_{\mathrm{cross~}p} \ll t_{\mathrm{formation~of~}q \bar q} .
  \end{equation}
  Taking the lowest-order, $q\bar q$ component gives the dipole model.

\item Breit frame.  In this frame, the hard interaction with the virtual
  photon occurs over a short time and distance scale.  The target is fast
  moving, time-dilated and Lorentz-contracted, so that the hard scattering
  occurs off a single quasi-free constituent of the target.  Thus the
  transverse distances and times obey
  \begin{equation}
     (r_T, t)_{\mathrm{hard~sc.}} \ll 
     (r_T, t)_{\mathrm{evolution~of~target}} .
  \end{equation}
  A factorization theorem is derived which relates the measured cross
  section to parton density functions (pdfs).

\end{enumerate}

The predictive power of the dipole model \cite{dipole} comes from the
observation that the transverse structure of the $q\bar q$ dipole is
frozen as it crosses the target.  The important concepts are that of the
light-front wave function of the virtual photon and the cross section for
the dipole to scatter off the target.  The picture is valid at the leading
logarithm level, and, by the use of the optical theorem, relations are found
between inclusive and diffractive cross sections.  The model is also
applied to real photon cross sections.  Its precise quantitative validity
is not clear.

The parton-density point of view gets its power from the observation that
just before the hard scattering, the struck quark in the target is
separated from the other constituents of the target and may be treated as
independent.  Important concepts are the number densities of quarks and
gluons {\em conditional} on the diffracted proton.  Also important are the
hard-scattering coefficients and the DGLAP evolution of the pdfs, which
are the same as in inclusive scattering.  The results are valid for the
whole leading power in $Q$, not just for the leading logarithms, and the
perturbative part is known to at least the first non-leading order.

Of course the cross sections are the same in the two frames.  However, the
intuition in the two frames is very different.  In particular, the time
sequence of the scattering is different in the two frames.  This is an
important symptom that there is non-trivial physics in relating the two
points of view.  In particular it implies that the struck parton (in the
pdf view), before it is struck, must have propagated over a {\em
space}-like distance, for otherwise the time-sequence would have been
preserved under a Lorentz transformation.

This indicates that the process is probing some interesting quantum
mechanical properties of QCD related to the EPR phenomenon and Bell's
theorem.  Although there are definite predictions from the theory, the
interpretation of phenomena in terms of a microscopic sequence of cause
and effect is frame (and observer) dependent.

The practitioners of the dipole picture gain information about
non-perturbative physics by treating dipole cross sections similarly to
hadronic cross sections, which gives a lot of useful, if only quantitative
information.  On the other hand, the practitioners of the pdf point of view
are agnostic about non-perturbative physics; but they are able to make
various non-trivial predictions from first principles that relate
different cross sections.

\section{Statement of factorization}

Hard-scattering factorization is valid \cite{diff.fact.proof} in
diffractive deep-inelastic scattering for the diffractive structure
functions $F_{2i}^{(D)}$ themselves, and for subprocesses like jet
production and heavy quark production.  That is, it is valid for any
process for which a factorization theorem is valid in inclusive
scattering; diffraction simply differs by a requirement in the target
fragmentation region.  The factorization theorem for these diffractive
processes states that the cross section differential in the relevant
variables is of the form
\begin{equation}
  \label{eq:fact}
  d\sigma = \sum_i \int d\xi f^{(D)}_i(\xi, x_\Pomeron, t; \mu)
                 d\hat\sigma_i ,
\end{equation}
up to corrections that are power suppressed in $Q$.  Here:
\begin{enumerate}
\item The index $i$ is the flavor of the struck parton.
\item The variable $\xi$ is the fractional light-front momentum of the
  struck parton relative to the ``Pomeron'', and $t$ and $x_\Pomeron$
  have their usual definitions.  Thus $\xi = k^+/(x_\Pomeron p^+)$,
  where $k$ is the momentum of the struck parton, and $p$ is the
  momentum of the target.
\item The hard-scattering coefficients $d\hat\sigma$ are perturbatively
  calculable and are the same as for the corresponding fully inclusive
  cross sections.  
\item The renormalization/factorization scale $\mu$ should be of order
  $Q$. 
\item The diffractive pdf $f^{(D)}_i$ is to be interpreted as the number
  density of partons conditional on the observation of a diffracted proton
  in the final state.
\item It obeys the standard DGLAP equation for its $\mu$ dependence with
  the same kernels as for the fully inclusive pdfs.
\item The result does not only apply in the diffractive region
  $x_\Pomeron \ll 1$, but also at larger values of $x_\Pomeron$.  The
  diffractive pdfs therefore are the same quantities as the extended
  fracture functions of Trentadue and Veneziano \cite{frac.fn}.
  Indeed any requirement may be applied in the target fragmentation
  region, relative to the target, and an obvious generalization of the
  theorem applies.
\end{enumerate}

The proof of the factorization formula also appears to be valid for
the {\em direct} photoproduction of jets, heavy quarks, etc.  However,
as we will see later, the factorization formula fails for
hadron-hadron scattering and therefore for {\em resolved}
photoproduction.  (This also implies that the fracture function method
fails for the same processes.)  Since the separation between direct
and resolved photoproduction is scheme-dependent beyond leading order,
the quantitative accuracy of factorization in photoproduction is not
completely clear.

Originally, the factorization formula was motivated by the model of
Ingelman and Schlein \cite{IS}, who used the common idea that
diffraction is due to exchange of a ``Pomeron''; they proposed that
diffractive hard scattering should be treated as being ordinary
hard scattering on a Pomeron target.  Associated with this is the
concept of parton densities in a Pomeron.  Unfortunately, Regge theory
in the form used by Ingelman and Schlein has not been derived from
QCD, and indeed is probably false; this is indicated experimentally by
the lack of universality of the Pomeron parameters between different
diffractive processes.  The factorization theorem that has actually
been proved, and is stated above, is somewhat different; it has
hard-scattering factorization but not Regge factorization.
Non-perturbative Regge properties of the process are in the
diffractive parton densities.

The diffractive parton densities have a simple interpretation in terms of
light-front creation and annihilation operators of partons.  If the
ordinary inclusive parton densities are defined as
\begin{equation}
  \label{eq:inclusive.pdf}
  x f^{\rm usual} = \int \frac{ d^2k_T }{ 2 (2\pi)^3 }
           \frac{ \langle p | a^\dagger_{x,k_T} a_{x,k_T} | p \rangle }{ \langle p | p \rangle } ,
\end{equation}
then the diffractive parton densities are
\begin{equation}
  \label{eq:diffractive.pdf}
  \frac{x}{x_\Pomeron} f^{(D)} = \sum_X \int \frac{ d^2k_T }{ 2 (2\pi)^3 }
           \frac{ \langle p | a^\dagger_{x,k_T} |Xp'\rangle
              \langle Xp'| a_{x,k_T} | p \rangle  }
           { \langle p | p \rangle }.
\end{equation}
Here the sum over intermediate states is restricted to those obeying
the correct diffractive condition.  The extra factor of $1/x_\Pomeron$
on the left-hand side of the definition of the diffractive pdf arises
because it is convenient to define $f^{(D)}$ as a number density in
the fractional momentum of the parton relative to the Pomeron rather
than relative to the proton.  The factors of $1/[2(2\pi)^3]$ are
associated with the normalization of the light-front operators.

{\em In these definitions, there are some important complications
  concerning UV renormalization and particularly concerning the correct
  gauge-invariant definition of the annihilation and creation operators
  that I will not go into here.}

These formulae can be expressed in terms of quark and gluon fields.  For
example the gauge-invariant definition of a diffractive quark density is
\begin{equation}
\fl
  \label{eq:diffractive.pdf.2}
  f^{(D)} = x_\Pomeron \sum_X \int \frac{ dy^- }{ 4\pi }
           e^{-i x_\Pomeron \beta p^+ y^-}
           \langle p | \bar\psi(0,y^-,0_T)  P(y^-)^\dagger |Xp'\rangle
           \langle Xp'| \gamma^+ P(0) \psi(0) | p \rangle .  
\end{equation}
Here $\psi$ is the ordinary Dirac field for the quark, and $P(y^-)$ is the
following path-ordered exponential of the gluon field:
\begin{equation}
  \label{eq:P}
  P(y^-) = {\cal P} \exp\left( -ig \int_{y^-}^\infty
                               d{y'}^- A^-(0,{y'}^-,0_T)
                        \right).
\end{equation}
The path-ordered exponential generates an eikonal line in the
Feynman-diagram representation of the parton density, and it can be
interpreted physically as the struck parton after the hard interaction in
the approximation of ignoring its deflection by final-state interactions.

\section{Why is factorization useful?}

\begin{enumerate}

\item Because of the universality of the (diffractive) pdfs between
  processes, and because of their DGLAP evolution, many predictions can be
  made from first principles, with the aid of perturbative hard-scattering
  calculations. 

\item It isolates the irreducible non-perturbative information in the
  parton densities alone.  

\item Hence any treatments of the non-perturbative part of the processes
  need only concern themselves with the diffractive pdfs alone; they do
  not need to make separate predictions for $F_2^{(D)}$, diffractive heavy
  quark production, diffractive jet production, etc.
  
\item The theorem is making non-trivial predictions about the dynamics of
  diffractive processes, in particular about the effect of final-state
  interactions.  The non-triviality of the results is illustrated by the
  fact that the interactions making the diffracted hadron occur later than
  the hard scattering and by the lack of factorization in diffractive
  hadron-hadron scattering.
  
\item Since the diffractive gluon density is much larger than the
  diffractive quark densities \cite{ACTW,H1.fit}, diffraction provides a
  gluon-rich target.

\item As mentioned in the previous section, there is an interesting
  space-time structure for the process.  There is an ambiguity as to what
  causes what.  This is likely to have interesting implications for
  understanding quantum mechanical states in non-perturbative QCD. 

\end{enumerate}

\begin{figure}
  \centering
  \includegraphics[width=0.7\textwidth]{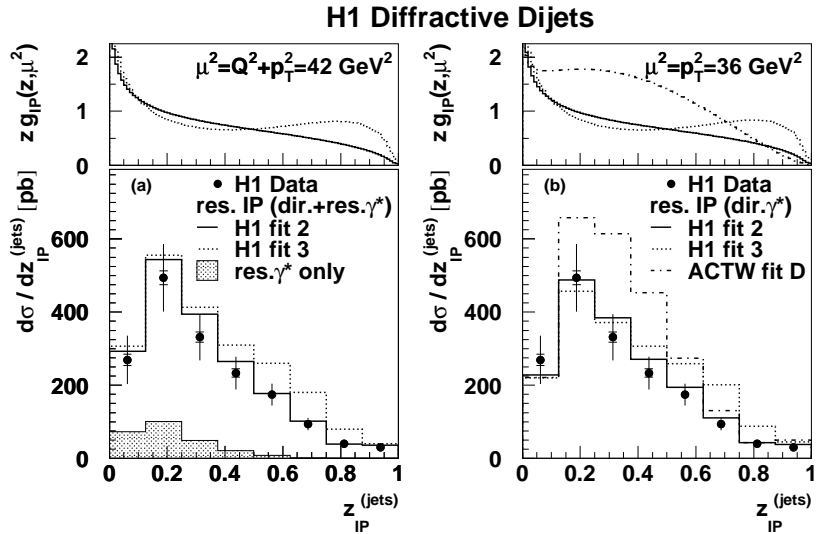}
  \caption{\label{fig:DDIS.jets} 
    Diffractive electroproduction of jets, from \cite{H1.dijets}.  The 
    variable $z_\Pomeron^{\rm (jets)}$ is the $+$ component of the
    dijet's momentum relative to $x_\Pomeron p^+$. 
  }
\end{figure}

\begin{figure}
  \centering
  \includegraphics[width=0.5\textwidth]{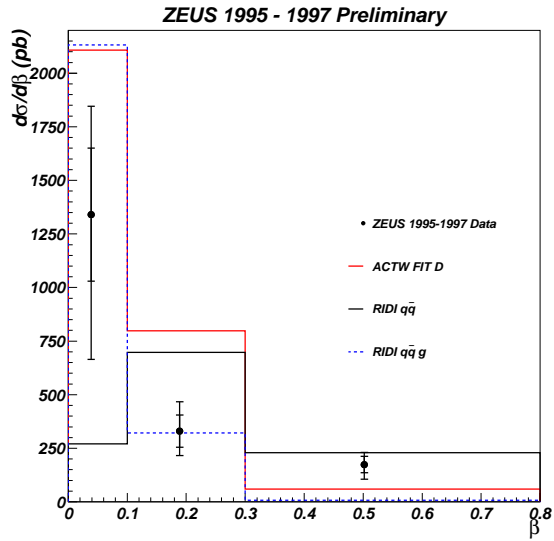}
  \caption{\label{fig:DDIS.charm} 
    Diffractive electroproduction of charm, from \cite{ZEUS.Dstar}.
  }
\end{figure}

Some of the predictions of the factorization theorem are shown in figures
\ref{fig:DDIS.jets} and \ref{fig:DDIS.charm}.  In these figures, recent
data are compared with predictions using fits of diffractive pdfs to other
data.  The H1 fits were obtained purely from diffractive $F_2^{(D)}$,
while the ACTW fits included also the earliest data on {\em direct}
photoproduction of jets.  The predictions work reasonably well,
and in particular strongly confirm the need for a large diffractive gluon 
density.

\section{Why is factorization valid \cite{diff.fact.proof}?}

\begin{figure}
  \centering
  \includegraphics[scale=0.45]{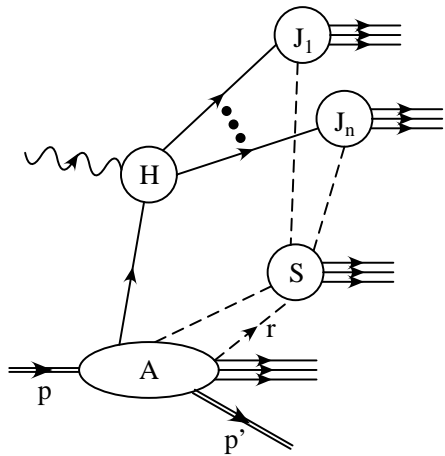}
\caption{Leading regions for the amplitude for diffractive DIS.}
\label{fig:regions}
\end{figure}

The leading regions for the amplitude for diffractive DIS are as shown
in figure \ref{fig:regions}.  There the hard subgraph $H$ contains
lines all of which are off-shell by order $Q^2$ and it results in one
or more jets going into the final state.  Connected to it by
essentially a single parton line is the target subgraph $A$, which
consists of lines collinear to the target, and includes the outgoing
diffracted proton.  Finally, there is a soft subgraph $S$, which
consists of lines that are of low energy with respect to both the jet
and target subgraphs; it includes the final-state interactions
responsible for the neutralization of the color of the partons
initiating the jets.

Regions not of this form are power suppressed.  Notably, situations in 
which extra partons are exchanged between the hard subgraph and either 
the target subgraph or the jet subgraphs are suppressed, because extra 
lines in the hard subgraph are then far off-shell.  Of course, if the
gluon density is too high, as in the saturation region, then the large
gluon density can overcome this suppression.

\begin{figure}
  \centering    
  \includegraphics[scale=0.45]{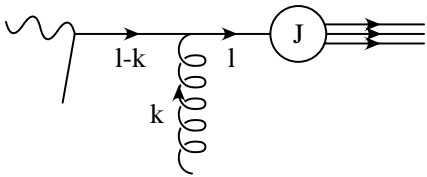}
\caption{Final-state interaction of jet.}
\label{fig:fsi}
\end{figure}

The factorization theorem now follows, essentially as in the case of
inclusive cross sections --- see \cite{diff.fact.proof}.  However
there are a number of complications to note:
\begin{enumerate}
\item Extra exchanges of collinear {\em longitudinally} polarized
  gluons between the hard subgraph and the collinear subgraphs are in
  fact allowed and unsuppressed, in some gauges, like the Feynman
  gauge.  After an appropriate approximation, we use Ward identities
  to show that these contributions give the path-ordered exponentials
  in the gauge-invariant definition of the parton densities.

\item This argument does not directly apply to all the final-state
  interactions, as I now illustrate with the aid of figure
  \ref{fig:fsi}.  In the most obvious region 
  of quasi-elastic scattering, the exchanged momentum has much smaller
  longitudinal momentum than transverse momentum: $|k^+k^-|\ll k_T^2$,
  and the approximations and the Ward identities simply do not apply.
  However, the only significant dependence on $k^+$ is in the
  propagator of the quark line emerging from the hard scattering:
  \begin{equation}
    \label{eq:fsi}
    \frac{1}{ (l-k)^2-m^2+i\epsilon }
    \simeq \frac{1}{ -2l^-k^+ + \mbox{transverse} +i\epsilon } .
  \end{equation}
  The pole in $k^+$ is in the upper-half plane, so that the contour of
  integration can be deformed away from the pole.  
  
  The limit to the contour deformation is when $k^+$ is so big that
  the dependence of other propagators on $k^+$ is non-negligible.
  Then the line $l-k$ is far off-shell, and we have a short-distance
  interaction, not a genuine final-state interaction, and the previous
  argument about the longitudinally polarized gluons and the Ward
  identities applies.
  
\item Even when there are more complicated interactions of the
  exchanged system and the jet, the singularities are only in the
  upper-half plane: this corresponds to the fact that the interactions
  are all in the final-state.  Thus the above argument generalizes.

\item A slight generalization of the above argument applies when the
  gluon is soft and obeys $k^+k^- \sim k_T^2$.  After the use of Ward
  identities, an application of unitarity is needed before
  factorization is obtained.

\end{enumerate}

Note that the argument about the final-state interactions is
non-perturbative and independent of the diffractive requirement on the
cross section.

\section{Exclusive diffraction}

In this section I consider exclusive diffractive processes, for
example deeply virtual Compton scattering (DVCS), which is $e+p \to e'
+ \gamma + p'$ at large $Q$, or inelastic meson production, $e + p \to e' + 
M + p'$, at large $Q$.  

\begin{figure}
  \centering    
  \includegraphics[scale=0.45]{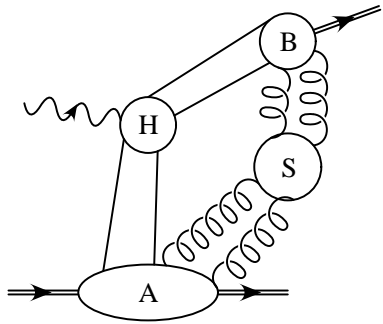}
\caption{Leading regions for deep-inelastic meson production.   The
  lower pair of quark lines may also be replaced by gluon lines.}
\label{fig:mesons}
\end{figure}

The same general idea as before applies \cite{mesons,Rad} to the
derivation of a factorization theorem, but now the derivation is
applied to the {\em amplitude}.  For deep-inelastic meson production
the leading regions have the form shown in figure \ref{fig:mesons}.
The resulting theorem is that the amplitude has the form
\begin{equation}
  \label{eq:fact.mesons}
  {\cal M} = \int d\xi \int dz f(\xi,x_\Pomeron,t,Q) H(\dots) \psi(z,Q),
\end{equation}
up to power-law corrections.  Here $H$ is the perturbatively
calculable hard-scattering coefficient, $\psi$ is the light-front wave
function of the meson, and $f$ is a skewed parton density.  Skewed
parton densities \cite{Rad,skewed.pdf} are defined just like ordinary
parton densities except that they use off-diagonal matrix elements of
the quark number operator instead of the expectation values.
Schematically, therefore, we have
\begin{equation}
  \label{eq:skewed.pdf}
  f = \mbox{normalization} \int d^2k_T 
           \langle p' | a^\dagger_{x,k_T} a_{x,k_T} | p \rangle .
\end{equation}
There are some varied conventions concerning the normalization factor.
All the same complications concerning final-state interactions, etc,
apply as for the factorization theorem for inclusive scattering.

Observe that the cross section is {\em quadratic}, not linear, in the
skewed parton density.

\section{Lack of factorization in diffractive hadron-hadron scattering}

Factorization does not hold for hard processes in diffractive
hadron-hadron scattering, contrary to the case in diffractive
deep-inelastic scattering.  The problem is that soft interactions
between the two hadrons, and their remnants, occur in both the initial
and final state.  Thus the contour-deformation argument used in
diffractive DIS no longer applies.  

A more difficult argument is needed to give factorization
\cite{HH.fact}, and this only applies to fully inclusive cross
sections.  Part of the argument involves a unitarity sum over both
beam remnants, and this fails in the diffractive case.  Even before
the discovery of QCD, it was known \cite{preQCD} that factorization
fails for diffractive hadron-hadron scattering.  In addition, there is
another potential mechanism the failure of factorization in QCD, the
so-called coherent Pomeron mechanism \cite{coherent.pomeron}.

\begin{figure}
\centering
  \includegraphics[width=0.4\textwidth]{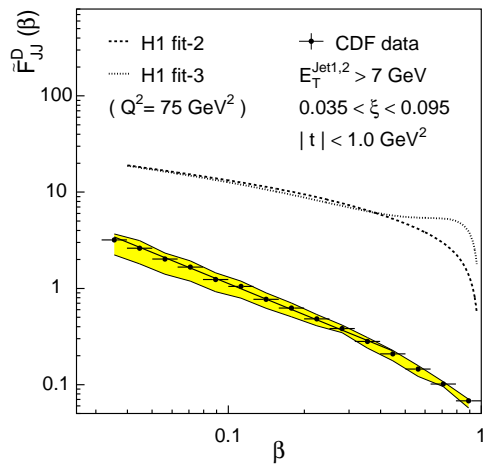}
  \caption{Comparison of CDF's diffractive dijet data with
    predictions on the basis of factorization and fits to HERA data
    \cite{H1.fit}, from \cite{CDF}.} 
  \label{fig:CDF}
\end{figure}

The theoretically predicted failure of factorization is in fact
observed experimentally.  The cross section for diffractive dijet
production at CDF \cite{CDF} is about an order of magnitude below what
factorization would predict given the diffractive parton densities
measured at HERA --- figure \ref{fig:CDF}.

\section{Conclusions}

\begin{itemize}
\item Factorization is proved for diffractive DIS and related
  processes.  It implies a separation of long- and short-distance
  phenomena, and makes non-trivial predictions.

\item The proof involves a non-trivial demonstration that final-state
  interactions do not affect factorization.

\item Factorization fails in diffractive hadron-hadron scattering.
  This is an experimentally verified consequence of QCD.

\item There are non-trivial tests of factorization in diffractive DIS,
  for example, the prediction of dijet cross sections.

\item It is interesting to fully understand the relation between the
  factorization/pdf approach and the dipole model.
\end{itemize}

\ack

I would like to thank DESY and the University of Hamburg for their
hospitality and the Alexander von Humboldt foundation for an award.
This work was supported in part by the U.S.\ Department of Energy
under grant number DE-FG02-90ER-40577.


\end{document}